\documentclass[12pt]{iopart}
\usepackage{graphicx}
\usepackage{epsfig,color}
\usepackage{hyperref}

\expandafter\let\csname equation*\endcsname\relax
\expandafter\let\csname endequation*\endcsname\relax
\usepackage{amsmath}
\usepackage{amsfonts}
\usepackage{amssymb}
\usepackage{cite}

\begin{document}

\title{Half a state, half an operator: a general formulation of stators}

\author{Erez Zohar}
\address{Max-Planck-Institut f\"ur Quantenoptik, Hans-Kopfermann-Stra\ss e 1, 85748 Garching, Germany.}

\begin{abstract}
Stators, which may be intuitively defined as "half states, half operators" are mathematical objects which act on two Hilbert spaces and utilize entanglement to create remote operations
and exchange information between two physical systems. In particular, they allow to induce  effective dynamics on one physical system by acting on the other one, given they have properly
been connected with the right stator. In this work, the concept of stators is generalized and formalized in a way that allows the utilization of stators for some physical
problems based on symmetry groups, and in particular digital quantum simulation.
\end{abstract}

\maketitle

\newpage

\section{Introduction}

Stators ("state operators") first introduced in \cite{Reznik2002}, are  somewhat peculiar mathematical objects, which may be intuitively defined as "half-states, half-operators";
more rigorously, they are defined on a product Hilbert space, $\mathcal{H}=\mathcal{H}_A \otimes \mathcal{H}_B$, and behave as operators acting on $\mathcal{H}_A$ and as states in $\mathcal{H}_B$.

Stators arose for the first time in the context of quantum information science. As an object defined on the product Hilbert space of two quantum systems, they were introduced and utilized for the study of remote operations, causality and locality implications on measurements and similar issues. In \cite{Reznik2002}, stators whose mathematical structure is based on finite Abelian groups - $\mathbb{Z}_N$ - were introduced, along with some possible quantum information applications. Other quantum information applications may be found in \cite{Reznik2002,Groisman2002,Groisman2005}.

 Another possible use of stators is for quantum simulation. One may consider one of the systems as an "ancillary system", which, once the stator is properly constructed, allows to create effective interactions between "physical systems" - i.e., two-body interactions may be used for the construction of dynamics involving three-body, four-body interactions and so on. However, for that one may need to use a more general formulation of stators, which is the aim of this work.

In this paper, we aim at generalizing the mathematical structure of a stator, in a way that allows, eventually, construct stators related to a group $G$, which is either finite or compact, Abelian or non-Abelian, using a Hilbert space structure based on the elements of $G$. This enables the use of stators for problems involving a given symmetry group.

We will start with a formal definition of stators, present some examples and then continue to a general prescription for constructing them.

\section{Definition of a stator}

\emph{Definition.} Let $\mathcal{H}_A$ and $\mathcal{H}_B$ be two Hilbert spaces, corresponding to the physical systems $A$ and $B$ respectively, and let $\mathcal{O}\left(\mathcal{H}\right)$
be the set of of operators acting on a Hilbert space $\mathcal{H}$. Then a stator $S$ is defined as an object of the form
\begin{equation}
S \in  \mathcal{O}\left(\mathcal{H}_A\right) \times \mathcal{H}_B.
\end{equation}
generated the action of a unitary operator
$U_{AB} \in  \mathcal{O}\left(\mathcal{H}_A \otimes \mathcal{H}_B \right)$ on a given state of the $B$ system, $\left|0_B\right\rangle$:
\begin{equation}
S = U_{AB}\left|0_B\right\rangle
\end{equation}

Suppose $\mathcal{H}_B$ is spanned by the basis $\left|i_B\right\rangle$; Then, one may define the Kraus operators
\begin{equation}
M_i = \left\langle i_B \right|U_{AB}\left|0_B\right\rangle,
\end{equation}
which satisfy
\begin{equation}
\underset{i}{\sum}M^{\dagger}_iM_i = \mathbf{1}_A,
\end{equation}
and the stator $S$ could be expanded as
\begin{equation}
S = \underset{i}{\sum}M_i \otimes \left|i_B\right\rangle
\end{equation}
Mathematically, $S$ is an isometry, mapping states in $\left|\psi_A\right\rangle \in \mathcal{H}_A$ to a tensor product $U_{AB}\left(\left|\psi_A\right\rangle \otimes \left|0_B\right\rangle\right) \in \mathcal{H}_A \times \mathcal{H}_B$.

Stators may be constructed such that they satisfy the eigenoperator relation
\begin{equation}
\Theta_B S = S \Theta_A
\end{equation}
where $\Theta_A,\Theta_B$ are operators acting on $\mathcal{H}_A$ and $\mathcal{H}_B$ respectively
\footnote{While here we focus on this particular type of eigenoperator relations, it is not the most general one, and one may consider others as well, e.g. $\Theta_B S = \Theta_A S$ etc.}; This can be utilized, for example, to obtain effective dynamics for system A
with the use of system B; suppose we have an initial state $\left|\psi_A\right\rangle$ of system $A$, which we want to evolve in time with some Hamiltonian $H_A$. This can be done using a
stator $S$ which satisfies
\begin{equation}
H_B S = S H_A
\end{equation}
for some $H_B$, since that implies (as the Hamiltonians $H_A,H_B$ are Hermitian operators)
\begin{equation}
e^{-i H_B t} S = S e^{-i H_A t}.
\end{equation}
This works as follows; Start with the initial state $\left|\psi_A\right\rangle \left|0_B\right\rangle$, create the stator and evolve system $B$ in time:
\begin{equation}
e^{-i H_B t} U_{AB} \left|\psi_A\right\rangle \left|0_B\right\rangle =  e^{-i H_B t} S \left|\psi_A\right\rangle
\end{equation}
using the eigenoperator relation we get
\begin{equation}
e^{-i H_B t} U_{AB} \left|\psi_A\right\rangle \left|0_B\right\rangle =  S e^{-i H_A t} \left|\psi_A\right\rangle = U_{AB} \left|0_B\right\rangle e^{-i H_A t} \left|\psi_A\right\rangle
\end{equation}
so, effectively, we can get the desired dynamics of a \emph{"physical" system} $A$ through interacting with the \emph{"ancillary" system} $B$ and letting it evolve in time:
\begin{equation}
e^{-i H_A t} \left|\psi_A\right\rangle = \left\langle 0_B\right| U^{\dagger}_{AB} e^{-i H_B t} U_{AB} \left|\psi_A\right\rangle \left|0_B\right\rangle
\end{equation}
This can be used, for example, in quantum simulation, when one, for various reasons, wishes to create  effective dynamics of a system $A$ by using an ancillary system $B$ \cite{ZNSim,Statsim}.

Matrix Product States (MPS) and Projected Entangles Pair States (PEPS) \cite{Verstraete2008,Orus2014} may also be seen as stators; such physical states are created by contracting
some ancillary degrees of freedom, from a tensor network whose basic ingredients are practically stators, whose state part is "physical" and operator part - auxiliary, which will be contracted in a special manner of the creation of a physical many body state. Such states may be parameterized in a way which guarantees symmetry -  invariance under transformations belonging to some
group - which is obtained after solving an equation on the single site level, connecting both types of degrees of freedom \cite{Schuch2010}. This equation is simply an eigenoperator relation.

Another appearance of a stator may be found in a similar context, when one wishes to create fermionic PEPS with a local gauge symmetry \cite{Zohar2015b,Zohar2016}. In this case, the local information of a many-body state is
encoded in a stator, with a fermionic (matter and virtual) part is described in terms of second quantized operators, as well as a gauge field part which is described by states.

\section{Examples of stators}
In the following section we shall address the of construction of stators: given the desired $\Theta_A$, what are the choices one can (and has to) make for the ancillary system $B$,
the operator $\Theta_B$ and the Kraus operators $M_i$ (or, equivalently, $U_{AB}$)? However, before turning to that, for a better intuition and understanding, let us consider some
particular examples.

\emph{Example 1 - spin $1/2$.} Consider the case in which both the systems $A$ and $B$ are spin-half particles, with the stator
\begin{equation}
S = \frac{1}{\sqrt{2}}\left(1_A \otimes \left|\uparrow_B\right\rangle + \sigma_A \otimes \left|\downarrow_B\right\rangle\right)
\end{equation}
where $\sigma$ is a Pauli operator in some given direction, $\sigma^2 = 1$. It is straightforward to verify that $\frac{1}{\sqrt{2}}1_A, \frac{1}{\sqrt{2}}\sigma_A$ form a complete set of Kraus operators;
this stator fulfills the eigenoperator relation
\begin{equation}
\sigma_{x,B} S = S \sigma_A
\end{equation}

\emph{Example 2. - $\mathbb{Z}_N$.} The previous example was, in fact, of a stator with a $\mathbb{Z}_2$ structure. This may be generalized to any $\mathbb{Z}_N$ group. Define the operators
$Q,P$ satisfying
\begin{equation}
P^{\dagger}P=Q^{\dagger}Q=P^N=Q^N=1
\end{equation}
as well as
\begin{equation}
P^{\dagger}QP=e^{i \frac{2\pi}{N}}Q
\end{equation}
for some given integer $N$. These act within an $N$ dimensional Hilbert space, for which one may choose, for example, the two bases $\left\{\left|m\right\rangle\right\}_{m=1}^{N}$
and $\left\{\left|\alpha\right\rangle\right\}_{\alpha=1}^{N}$, such that
\begin{equation}
P\left|m\right\rangle=e^{i \frac{2\pi m}{N}}\left|m\right\rangle \quad ; \quad
Q\left|\alpha\right\rangle=e^{i \frac{2\pi\alpha}{N}}\left|\alpha\right\rangle \quad ; \quad
Q\left|m\right\rangle=\left|m-1\right\rangle \quad ; \quad
P\left|\alpha\right\rangle=\left|\alpha +1\right\rangle \\
\label{Znalg}
\end{equation}
and $\left\langle m | \alpha \right\rangle = \frac{1}{\sqrt{N}}e^{i \frac{2\pi m \alpha}{N}}$.
Then one may define, for example, for such $A,B$ systems the stator,
\begin{equation}
S = e^{-i \frac{N}{2\pi} \log\left(Q_A\right)\log\left(P_B\right)}\left|\alpha = 0_B\right\rangle = \frac{1}{\sqrt{N}}\overset{N}{\underset{m=1}{\sum}}Q_A^{m}\otimes\left|m_B\right\rangle
\end{equation}
satisfying
$Q_B S = S Q_A$. The case $N=2$ corresponds to the previous example, if $\sigma=\sigma_x(=Q)$.
Using (\ref{Znalg}), let us convert the above $S$ to the basis of $Q$ eigenstates, and obtain simply
\begin{equation}
S = \overset{N}{\underset{\alpha=1}{\sum}}\left|\alpha_A\right\rangle\left\langle\alpha_A\right|\otimes\left|\alpha_B\right\rangle
\end{equation}
from which one can easily see that the operators $M_{\alpha}=\left|\alpha\right\rangle\left\langle\alpha\right|$, as a complete set of projection operators,
form a complete set of Kraus operators as well.

\emph{Example 3. - General group $G$.} Finally let us generalize the previous case as well, for any group $G$ which is either a compact Lie or a finite group, either Abelian or non-Abelian.
Denote the group elements by $g\in G$, and introduce the "group elements basis", $\left|g\right\rangle$ \cite{Zohar2015,Zohar2016}. These  are eigenstates of the unitary operators $U^{j}_{mn}$ -  matrices of operators, belonging to the $j$ irreducible representation of $G$, with the eigenvalue equation
\begin{equation}
U^{j}_{mn} \left|g\right\rangle = D^{j}_{mn}\left(g\right)\left|g\right\rangle
\label{Udiag}
\end{equation}
where $D^{j}_{mn}\left(g\right)$ is the unitary \emph{Wigner matrix} \cite{Landau1981,Rose1995} representing $g$ in the $j$ irreducible representation of $G$,
with dimension $\dim\left(j\right)$.

The stator from the previous example may be generalized to
\begin{equation}
S = \int dg \left|g_A\right\rangle \left\langle g_A\right| \otimes \left|g_B\right\rangle
\label{gelsta}
\end{equation}
satisfying
\begin{equation}
\left(U^{j}_{mn}\right)_B S = S \left(U^{j}_{mn}\right)_A
\end{equation}
(note that for $\mathbb{Z}_N$, this correspond to the $Q$ action). This type of a stator shall be called \emph{Group element stator}. All the previous examples are of this type.

On the other hand, one may express the states of this "group Hilbert space"  in another basis too, which we call \emph{Representation basis} \cite{Zohar2015,Zohar2016} . This basis consists of the states
$\left|jmn\right\rangle$, where $j$ is a number (or set of numbers) labeling the irreducible representation, and $m,n$ are numbers (or sets of numbers) denoting the eigenvalues of operators
which are block-diagonal in the irreducible representations, mutually commuting among themselves, corresponding to the left and right degrees of freedom of the group - as $G$ may be, in general,
non-Abelian, and thus acting with group transformations on the left or on the right results, in general, in different outcomes. In $SU(2)$, for example, the irreducible representations
are labeled by the total angular momentum $j$, and the states may be finally described by the eigenvalue of one component of the angular momentum - usually the $z$ component. One has
two sets of angular momentum components, left - $L_i$ and right - $R_i$ (which correspond, in a mechanical picture, to the angular momentum of a rigid body in the space and body frames, and thus
commute with one another \cite{Landau1981}), and then,
\begin{equation}
\begin{aligned}
\mathbf{J}^2\left|jmn\right\rangle &= j\left(j+1\right)\left|jmn\right\rangle\\
L_z\left|jmn\right\rangle &= m\left|jmn\right\rangle\\
R_z\left|jmn\right\rangle &= n\left|jmn\right\rangle
\end{aligned}
\end{equation}
In $SU(3)$, on the other hand, the representations $j$ will have some multiplicity and thus $j$ has to consist of more than one number; Besides that, $m,n$ should both be a set of two
numbers, corresponding to the two mutually commuting operators, block-diagonal in the representation one may use there - the isospin and hypercharge, in particle physics terms.

The relation between these two bases is given in the following "Fourier transform",
\begin{equation}
\left\langle g | jmn \right\rangle = \sqrt{\frac{\dim\left(j\right)}{\left|G\right|}}D^{j}_{mn}\left(g\right)
\end{equation}
where $\left|G\right|$ is the order of the group.

In this basis, one may obtain using the Clebsch-Gordan series, that \cite{Zohar2015}
\begin{equation}
U^{j}_{mm'} =\underset{J,M,K,N}{\sum}\sqrt{\frac{\dim\left(J\right)}{\dim\left(K\right)}}\left\langle J M j m | K N \right\rangle\left\langle K N' | J M' j m' \right\rangle
\left| K N N' \right\rangle \left\langle J M M'\right|
\label{Urep}
\end{equation}
and the group element stator will thus have the form
\begin{equation}
S = \underset{j,m,m'}{\sum}\sqrt{\frac{\dim\left(j\right)}{\left|G\right|}}\left(U^j_{mm'}\right)^{\dagger}_A \otimes \left|j m m'_B\right\rangle
\end{equation}
in the representation basis.

\section{General construction of stators}
After having considered several examples for stators, which all  generalize to the group element stator given in example 3, we can turn on to a more general theory of stators,  allowing us to construct a stator for a given problem. That is, we wish to solve the following problem:

\emph{Given a Hilbert space $\mathcal{H}_A$, whose dimension is $N_A$, and an operator $\Theta_A \in \mathcal{O}\left(\mathcal{H}_A\right)$ (which will either be Hermitian, or an analytical function of an Hermitian operator),
what are the $\mathcal{H}_B$ and $\Theta_B \in \mathcal{O}\left(\mathcal{H}_B\right)$ required for the construction of a stator
$S \in \mathcal{O}\left(\mathcal{H}_A\right) \times \mathcal{H}_B$ satisfying the eigenoperator relation $\Theta_B S = S \Theta_A$?}

$\Theta_A$ may be diagonlized and brought to the diagonal form $\Lambda = \text{diag}\left(\lambda_1,...,\lambda_{N_A}\right)$, by some unitary operator $V$,
\begin{equation}
\Theta_A = V \Lambda V^{\dagger}
\end{equation}

We will denote the basis elements of (the yet unknown) $\mathcal{H}_B$ by $\left|\tilde i\right\rangle$. The matrix elements of (the yet unknown as well) $\Theta_B$ will be denoted by
\begin{equation}
R_{ij} = \left\langle \tilde i \right| \Theta_B \left| \tilde j \right\rangle.
\end{equation}
This matrix may be diagonalized by some unitary matrix $W$,
\begin{equation}
R = W \Omega W^{\dagger}
\end{equation}
with $\Omega =  \text{diag}\left(\omega_1,...,\omega_{N_B}\right)$

The stator will take the general form
\begin{equation}
S = \overset{N_B}{\underset{i=1}{\sum}}M_i \otimes \left|\tilde i\right\rangle
\end{equation}
Acting on it with $\Theta_A$ and $\Theta_B$ and demanding the eigenoperator relation leads to an operator equation in $\mathcal{H}_A$,
\begin{equation}
M_i \Theta_A = \underset{j}{\sum}R_{ij} M_j , \forall i
\end{equation}

Define the operators
\begin{equation}
N_i = M_i V
\end{equation}
as well as
\begin{equation}
K_i = \underset{j}{\sum}\left(W^{\dagger}\right)_{ij}N_j.
\end{equation}
Note that while the $N_i$ operators are obtained by the product of two operators in $\mathcal{H}_A$, the $K_i$ operators are merely linear combinations of the $N_i$ opeartors,
as $W_{ij}$ is a matrix of numbers and not an operator in $\mathcal{O}\left(\mathcal{H}_A\right)$. Both these sets of operators have to be Kraus operators, as a direct consequence of $M_i$ forming such a set,
and of $V$ being a unitary operator and $W_{ij}$ a unitary matrix.

One obtains then a simple equation for the $K_i$ operators,
\begin{equation}
K_i \Lambda = \omega_i K_i ,\forall i
\end{equation}
(no summation on $i$). That immediately implies that the minimal choice of $N_B$ is $N_B=N_A$. As we wish our construction to be minimal, we will take $N_B=N_A\equiv N$,
as well as $\mathcal{H}_B \simeq \mathcal{H}_A$. Furthermore, we get that $\Theta_A$ and $\Theta_B$ have the same spectrum, i.e. $\Lambda = \Omega$ (or, at least, they are equal up to
change of rows and columns). Finally, we get that a general solution for $K_i$ admits the form
\begin{equation}
K_i = \underset{\alpha}{\sum}\kappa_{\alpha i}\left|\alpha\right\rangle\left\langle i \right|
\end{equation}
Since the $K_i$ operators are Kraus operators, one obtains that
\begin{equation}
\underset{i}{\sum}\kappa_{\alpha i} \overline{\kappa}_{\beta i} = \delta_{\alpha \beta}
\end{equation}
i.e., $\kappa$ is a unitary matrix. The most trivial choice for $K_i$ would be projection operators, $K_i = \left|i\right\rangle\left\langle i\right|$, but we will keep our construction general.

Then one has to do the following choices:
\begin{enumerate}
  \item Choose $\mathcal{H}_B$, as long as its dimension is $N$, and the preferred basis $\left|\tilde i\right\rangle$ to work in.
  \item Although $\Omega$ is already determined since $\Theta_A$, $\Theta_B$ must have the same spectrum, choose the unitary $W_{ij}$ in order to obtain $R_{ij}$, which determines $\Theta_B$.
  \item Upon the choice of $K_i$ and $W_{ij}$, obtain $N_i$ and then $M_i$.
\end{enumerate}
Note that the $M_i$ operators depend on $V$, but not on $\Lambda$, while $\Theta_B$ depends on $\Lambda$ but not on $V$;
 that means that the stator depends on the eigenvectors of $\Theta_A$, rather than on its spectrum, and thus one may use the
same stator for a set of mutually commuting (and thus mutually diagonalizable) operators. Such operators differ only in their eigenvalues,
thus they will result in a different $\Theta_B$.

As an example, let us work out a simple, but rather general case, in which we deal with a group $G$ and its states $\left|g\right\rangle$. We wish to have $\Theta_A = U^{j}_{mn}$.
First we work in the group element basis, where $\Theta_A$ is diagonal ($V_{gh}=\delta_{gh}$, see eq. (\ref{Udiag})), and thus $M_g=K_g$. We make the simplest choices,
 $K_g= \left|g\right\rangle\left\langle g \right|$, and $W_{gh} = \delta_{gh}$, which imply that $\Theta_B = U^{j}_{mn}$, $N_g=K_g$ and thus
 $M_g= \left|g\right\rangle\left\langle g \right|$. Altogether we obtain the group element stator
introduced in eq. (\ref{gelsta}).

We may also calculate the same stator in the representation basis. There, $\Theta_A$ is not diagonal (see eq. (\ref{Urep})).
Using
\begin{equation}
\left\langle K N N' \right| U^{j}_{mm'} \left|J M M'\right\rangle = \int dg \left\langle K N N' | g \right\rangle \left\langle g | J M M' \right \rangle D^{j}_{mm'} \left(g\right)
\end{equation}
we get the diagonalizing matrix
\begin{equation}
V = \int dg \underset{jmm'}{\sum}\left\langle jmm' | g \right\rangle \left|jmm'\right\rangle\left\langle g \right|
= \int dg \underset{jmm'}{\sum}\sqrt{\frac{\dim\left(j\right)}{\left|G\right|}}\overline{D}^{j}_{mm'}\left(g\right) \left|jmm'\right\rangle\left\langle g \right|
\end{equation}
(note that this could be written as the identity operator using the completeness relations in both group element and representation bases, but we shall not do that as we wish it to act as a change of basis matrix, connecting the two bases).
Thus, the $M_{jmm'}$ in $S=\underset{jmm'}{\sum}M_{jmm'}\otimes\left|jmm'\right\rangle$ will be given by
\begin{equation}
\begin{aligned}
M_{jmm'}&=N_{jmm'}V^{\dagger} =
N_{jmm'} \int dg \underset{KNN'}{\sum}\sqrt{\frac{\dim\left(j\right)}{\left|G\right|}}D^{K}_{NN'}\left(g\right)\left|g\right\rangle\left\langle KNN'\right| \\&=
\int dg \int dh \underset{KNN'}{\sum}W_{jmm',h}\sqrt{\frac{\dim\left(j\right)}{\left|G\right|}}D^{K}_{NN'}\left(g\right)K_h\left|g\right\rangle\left\langle KNN'\right|
\end{aligned}
\end{equation}

In order to get $\Theta_B=\Theta_A$, we choose $W_{jmm',g}=V_{jmm',g}$ and $K_g = \left|g\right\rangle\left\langle g\right|$, and obtain
\begin{equation}
M_{jmm'}=\int dg \underset{JMM',KNN'}{\sum}\sqrt{\frac{\dim\left(j\right)\dim\left(J\right)\dim\left(K\right)}{\left|G\right|^3}}
\overline{D}^{j}_{mm'}\left(g\right) \overline{D}^{J}_{MM'}\left(g\right) D^{K}_{NN'}\left(g\right) \left|JMM'\right\rangle\left\langle KNN'\right|
\end{equation}
this is simplified using the Clebsch-Gordan series \cite{Rose1995},
\begin{equation}
\begin{aligned}
M_{jmm'}=\int dg \underset{JMM',KNN',IPP'}{\sum}\sqrt{\frac{\dim\left(j\right)\dim\left(J\right)\dim\left(K\right)}{\left|G\right|^3}}
 \overline{D}^{I}_{PP'}\left(g\right) D^{K}_{NN'}\left(g\right) \\ \times \overline{\left\langle JMjm | IP \right\rangle}
\overline{\left\langle IP' | JM'jm' \right\rangle}\left|JMM'\right\rangle\left\langle KNN'\right|
\end{aligned}
\end{equation}
and further by the great orthogonality theorem,
\begin{equation}
\begin{aligned}
M_{jmm'}&=\int dg \underset{JMM',KNN'}{\sum}\sqrt{\frac{\dim\left(j\right)}{\left|G\right|}} \sqrt{\frac{\dim\left(J\right)}{\dim\left(K\right)}}
  \overline{\left\langle JMjm | KN \right\rangle}
\overline{\left\langle KN' | JM'jm' \right\rangle}\left|JMM'\right\rangle\left\langle KNN'\right| \\&=
\sqrt{\frac{\dim\left(j\right)}{\left|G\right|}}\left(U^{j}_{mm'}\right)^{\dagger}
\end{aligned}
\end{equation}
as we found by a direct change of basis in the previous section.

\section{Using stators to implement many body interactions}

Stators may be useful, in particular, for connecting several parties for a many-body interaction. Consider, for example, a case in which we wish to induce, effectively, an
interaction among several physical systems, $\left\{\mathcal{H}_A^k\right\}_{k=1}^\mathcal{N}$, of the form $\Theta = \underset{k}{\prod}\Theta_A^k \in \mathcal{O}\left(\underset{k}{\bigotimes}\mathcal{H}_A^k\right)$, where $\Theta_A^k \in \mathcal{O}\left(\mathcal{H}_A^k\right)$. We wish to use an ancillay system B to introduce a "global stator",
$S$, such that
\begin{equation}
\Theta_B S = S \Theta_A
\label{eigopglob}
\end{equation}

This may be achieved by using a set of unitary operators $U_{AB}^k \in \mathcal{O}\left(\mathcal{H}^k_A \otimes \mathcal{H}_B \right)$:
We define the stator as
\begin{equation}
S=\underset{k}{\prod}U^k_{AB}\left|0_B\right\rangle
\label{globs}
\end{equation}
with a set of Kraus Operators
\begin{equation}
M_i = \left\langle i_B \right|\underset{k}{\prod}U^k_{AB}\left|0_B\right\rangle,
\end{equation}

Then, demanding the eigenoperator relation (\ref{eigopglob}) leads to similar arguments as in the simple case. However, if the physical Hilbert spaces and the operators
$\Theta^k_A$ are similar, it is easy to build the stator $S$ from simple stators: in this case, define a stator $S^k \in \mathcal{O}\left(\mathcal{H}_A^k\right) \times \mathcal{H}_B$
as
\begin{equation}
S^k = U^k_{AB}\left|0_B\right\rangle
\end{equation}
such that
\begin{equation}
\Theta_B S^k = S^k \Theta_A^k
\end{equation}

Construct $S^k$ with the prescription given above, but choose a basis of $\mathcal{H}_B$ such that it will be a stator created with
\begin{equation}
U^k_{AB} = \underset{m}{\sum}u^k_m \otimes \left|m_B\right\rangle\left\langle m_B\right|
\label{Uproj}
\end{equation}
where $u^k_m \in \mathcal{O}\left(\mathcal{H}_A^k\right)$, which we define as a \emph{projective stator}.
Then, it is straightforward to see that the stator defined by (\ref{globs}) with (\ref{Uproj}) satisfies the desired eigenoperator relation (\ref{eigopglob}).

This procedure may be useful for the creation of digital quantum simulations of models which involve many-body interactions, such as plaquette interactions in lattice gauge theories. These are four-body interactions, which, in several previous quantum simulation proposals \cite{AngMom,NA,Topological,Zohar2015a} are realized using perturbative methods. With the introduction of stators, however, these interaction terms can be implemented by using a sequence of two-body interactions with an ancillary system \cite{Statsim,ZNSim} which improves the feasibility of simulation by avoiding the use of perturbative, effective interaction terms.

\section{Summary}
In this paper, the concept of stators, first introduced in \cite{Reznik2002}, was generalized and formalized. We have given a definition of a stator, some examples and a quantitative
study of stators, their construction and some important properties.

Besides their possible uses in quantum information \cite{Reznik2002}, stators may be of great help for studies of many body quantum systems, either for a theoretical approach involving tensor network states, or for the construction of quantum simulators, for many body systems with at symmetry group.
For example, they can be utilized for the construction of digital
quantum simulations of lattice gauge theories, where they are utilized to tailor the four-body plaquette interactions out of a sequence of two-body interactions with ancillary degrees of freedom \cite{Statsim,ZNSim}.

\section*{Acknowledgements}
The author would like to thank Alessandro Farace, Benni Reznik and J. Ignacio Cirac for helpful discussions, and to acknowledge the support of the Alexander-von-Humboldt foundation.

\bibliography{ref}

\end{document}